# Second-harmonic imaging of plasmonic Pancharatnam-Berry phase metasurfaces coupled to monolayers of WS$_2$

Florian Spreyer[1], Ruizhe Zhao[2], Lingling Huang[2], and Thomas Zentgraf[1]

[1]Department of Physics, Paderborn University, Warburger Straße 100, 33098 Paderborn, Germany

[2]School of Optics and Photonics, Beijing Institute of Technology, Beijing 100081, China

**Abstract:**

The nonlinear processes of frequency conversion like second harmonic generation (SHG) usually obey certain selection rules, resulting from the preservation of different kind of physical quantities, e.g., the angular momentum. For SHG created by monolayer of transition-metal dichalcogenides (TMDCs) like WS$_2$, the valley-exciton locked selection rule predicts an SHG signal in the cross-polarization state. By combining plasmonic nanostructures with a monolayer of TMDC, a hybrid metasurface is realized which affects this nonlinear process due to an additional polarization conversion process. Here, we observe that the plasmonic metasurface modifies the light-matter interaction with the TMDC resulting in an SHG signal that is co-polarized with respect to the incident field, which is usually forbidden for solely monolayers of TMDC. We fabricate such hybrid metasurfaces by placing plasmonic nanorods on top of a monolayer WS$_2$ and study the valley-exciton locked SHG emission from such system for different parameters, such as wavelength and polarization. Furthermore, we show the potential of the hybrid metasurface for tailoring nonlinear processes by adding additional phase information to the SHG signal using the Pancharatnam-Berry phase effect. This allows direct tailoring of the SHG emission to the far-field.

**Keywords:** second-harmonic generation, 2D materials, transition metal dichalcogenide monolayer, metasurface, plasmonics

## 1. Introduction

In the past years, two-dimensional layered materials have been studied more intensely since the discovery of the properties of graphene [1]. Graphene provides promising and unique properties like high mechanical resilience and shows great potential for various applications. However, due to the lack of a bandgap in graphene, its usability for optical applications is limited. Therefore, other materials like monolayers of transition metal dichalcogenides were proposed as a promising alternative [2].

By down-scaling TMDCs to a monolayer (1L-TMDC), the indirect bandgap of the bulk material changes to a direct bandgap [3] with bandgap energies typically ranging from approximately 1-2 eV [4]. Recently, large-scale single-crystalline flakes of TMDCs have been fabricated by chemical vapor deposition, which makes them easier to characterize by optical measurements like Raman or photoluminescence spectroscopy [5] and at the same time simplify the combination with other material systems. A unique characteristic optical feature of a 1L-TMD is the relatively high absorbance of light which is greater than 15% for single crystalline monolayer with less than 1 nm in thickness [6]. The comparatively large light-matter interaction of a monolayer opens up the possibility for applications like ultra-thin photodetectors, transistors, diodes and solar panels [7-9].

The bandgap of 1L-TMDCs in the visible to near-infrared range lately draws attention for applications in plasmonics. To enhance and modify the light-matter interaction, plasmonic nanostructures with their large scattering cross-section and strong near-field enhancement seem to be suitable candidates [10-13]. The increased interaction might also increase nonlinear optical effects and provide access to higher nonlinearities in ultrathin systems. Recently, first experiments have studied the nonlinear responses of the combination of 1L-TMDCs with plasmonic nanostructures that were combined to hybrid metasurfaces [14-17]. It was shown, that hybrid metasurfaces can be used to modify the nonlinear response in diverse ways. The strong near-field of plasmonic metasurfaces can enhance the nonlinear response due to the field overlap with the 1L-TMDCs. The strong coupling in such hybrid metasurfaces can also be exploited to address different valley excitons, which are connected to the nonlinear response directly [14].

Metasurfaces consisting of periodic arrays of plasmonic nanostructures with subwavelength sizes are capable of manipulating light on small scales in different ways [18]. In particular, spatial tailoring of nonlinear processes in amplitude and phase demonstrated the potential for ultrathin nonlinear optical elements [19, 20]. In the linear regime, the influence of plasmonic resonances of nanoholes below 1L-TMDC has been investigated. It was shown that plasmon modes at ~2 eV exhibit changes in the linear scattering spectra [21]. Other recent work studied quenching effects by uniformly shaped nanocone arrays with various periodicities coupled to monolayer 1L-TMDC. It was found that the PL intensity of 1L-TMDC was either increased or decreased by changing the spacing between plasmonic nanocones [22]. Although coupling effects between plasmonic nanostructures and 1L-TMDCs are reported for linear optical effects, the nonlinear coupling in hybrid metasurfaces is still largely unexplored.

Here, we experimentally demonstrate that plasmonic metasurfaces can alter the second harmonic generation of $WS_2$ monolayers in polarization and phase. In particular, we show that the interaction with the plasmonic nanostructures results in the emission of a circular polarization state of the second harmonic light that for pure $WS_2$ is forbidden by the symmetry selection rules for nonlinear processes. By imaging the second harmonic light from the patterned metasurfaces for different polarization configurations in real space, we visualize spatially the interaction effect and the modification of the second harmonic generation (SHG).

Our hybrid metasurface consists of Au-nanorods with $C_2$ rotational symmetry that are arranged in a square lattice on top of a WS$_2$ monolayer (1L-WS$_2$) placed on a quartz glass substrate (Figure 1A). The inversion symmetry of the nanorods has the advantage that the plasmonic structures themselves show a neglectable SHG signal and can act solely as a facilitator for the SHG of the WS$_2$. The nanorods are designed to be resonant at a fundamental wavelength of 1240 nm, which is half of the bandgap energy of the 1L-WS$_2$ corresponding to a wavelength of 620 nm (Figure 1B). In such way, the localized surface plasmon polariton resonances of the Au-nanorods can facilitate the coupling by a two-photon process to the WS$_2$ to lead to enhanced SHG. To approach the best overlap of the nanorods' resonance wavelengths with half of the bandgap energy of the 1L-WS$_2$, we fabricate three metasurfaces with different nanorod lengths [23]. In this way, the slightly modified coupling between the nanorods and the 1L-WS$_2$ is obtained and fabrication tolerances can be compensated. The metasurfaces are designed with spatially varying orientation angles of the nanoantennas ($\Delta\theta = \pm 70°$ between horizontal neighbored nanorods) in a square lattice with identical periods of $p = 400$ nm in both directions.

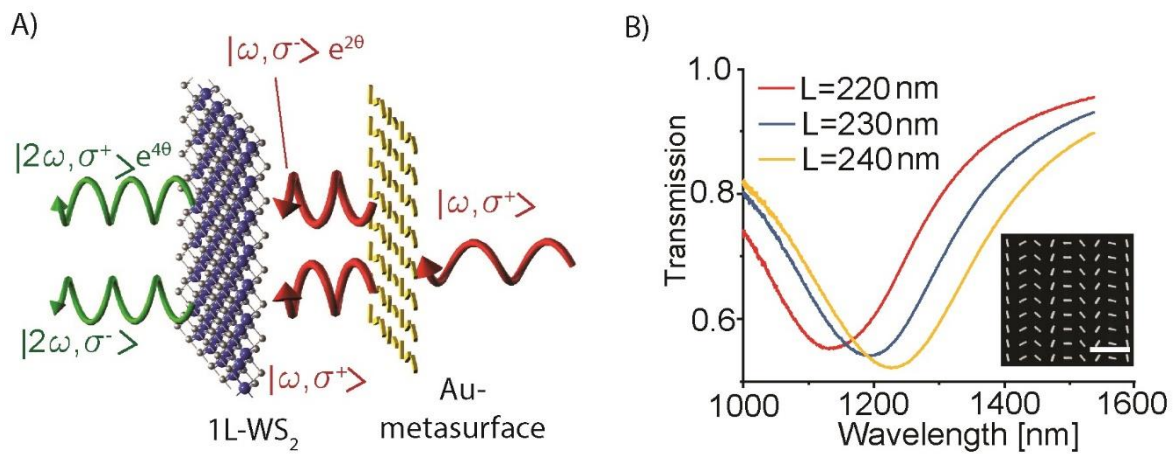

**Figure 1: Hybrid metasurface for nonlinear frequency conversion.** (A) Schematic illustration of the hybrid metasurface consisting of plasmonic Au-nanorods that are placed directly on top of monolayer WS$_2$. An incident laser beam with a frequency $\omega$ and right circular polarization $\sigma^+$ (RCP) interacts with the metasurface. The scattered field by the nanorods can be decomposed into the superposition of both circular polarization states $\sigma^+$ (RCP) and $\sigma^-$ (LCP), which couple to the WS$_2$ and result in a second harmonic generation (SHG). Note that only the $\sigma^-$ state of the scattered fundamental light by the nanorods carries an additional phase of $2\theta$ due to the orientation of the nanorods. After the interaction with the WS$_2$ the phase of the $\sigma^+$ SHG signal doubles to $4\theta$. (B) Transmission spectra of hybrid metasurfaces with three different antenna lengths and unpolarized light. The resonance wavelengths of the nanoantennas can be tuned to match the resonance wavelength of the two-photon absorption process of 1L-WS$_2$ at 1240 nm by optimizing the antenna length. The inset shows a scanning electron microscopy (SEM) image of the Au-metasurface with antenna lengths of 220 nm (scale bar equals 1 μm).

## 2. Fabrication & linear optical characterization

For our study, we choose monolayers of WS$_2$ as a suitable 1L-TMD due to the bandgap in the visible region at around 620 nm (2 eV), which can give strong SHG emission for excitation with laser pulses at a fundamental wavelength in the near-infrared region. The monolayer WS$_2$ flakes are grown via chemical vapor deposition and transferred to a quartz glass substrate. Next, we fabricate gold nanorod antennas on top of the WS$_2$ flakes by standard electron beam lithography. After the electron beam lithography and development of the PMMA photoresist mask, the sample is placed into an electron beam evaporator for the gold deposition. In the last step, the PMMA mask is removed by a lift-off process while the nanorod antennas remain on the surface (see inset of Figure 1B).

For the optical characterization, we first measure the photoluminescence (PL) signal of the WS$_2$ flakes by direct excitation with a wavelength of 532 nm. The emitted PL-light (with and without the metasurface) is measured in transmission, filtered by a bandpass filter and analyzed by a camera and a spectrometer. We observe a strong PL-signal at 614 nm that confirms the monolayer thickness of the flakes and shows the arrangement of the flakes on the quartz glass surface (Figures 2 A-D).

From our measurements, we observe that the PL intensity in the areas around the Au-metasurface is higher than the PL light coming directly from the metasurface covered areas. Here, the lower PL intensity from the metasurface areas is a result of the partially reflected and absorbed incident light by the nanorod antennas at 532 nm. In addition, the lower PL intensity for the direct excitation of the excitons can also be a result of quenching effects due to the nearby metal structures. Previous studies have shown such quenching effects for plasmonic nanostructure arrays [22]. However, the PL intensity difference allows us a clear identification of the spatial location of the metasurfaces and the 1L-WS$_2$ flakes.

In addition to the PL images, we analyze the spectral distribution of the PL light (Figure 2E). For the spectrally resolved PL, we do observe a small shift of the PL resonance peak in the spectrum if the nanorod antennas are placed on the top of the 1L-WS$_2$ flake. The shift in the PL spectrum is only visible for the nanoantennas with the length of L = 220 nm. However, we like to note that the shift is less than 1 nm and only observable in this particular antenna pattern. Based on other samples that have been fabricated previously but where not used for the further measurements, we did not observe any shift within our measurement tolerance. We anticipate that this observed shift arises by some experimental uncertainties. The peak positions of the other PL curves of the hybrid metasurfaces are at the same positions as the PL peak of bare 1L-WS$_2$. Therefore, we conclude that the bandgap energy has not changed significantly by the fabrication process of the plasmonic nanorods. Furthermore, as mentioned before, the PL intensity is not influenced by the nanorod dimensions giving a similar signal strength for all three metasurfaces. As the nanorod antennas are designed to be resonant in the near-infrared at 1240 nm, there is only a weak coupling effect for the excitation at 532 nm and the emission at 620 nm. This is supported by the lack of resonances in the transmission spectra in the visible regime, which shows no significant resonance at either 532 nm or 620 nm (see supplementary material).

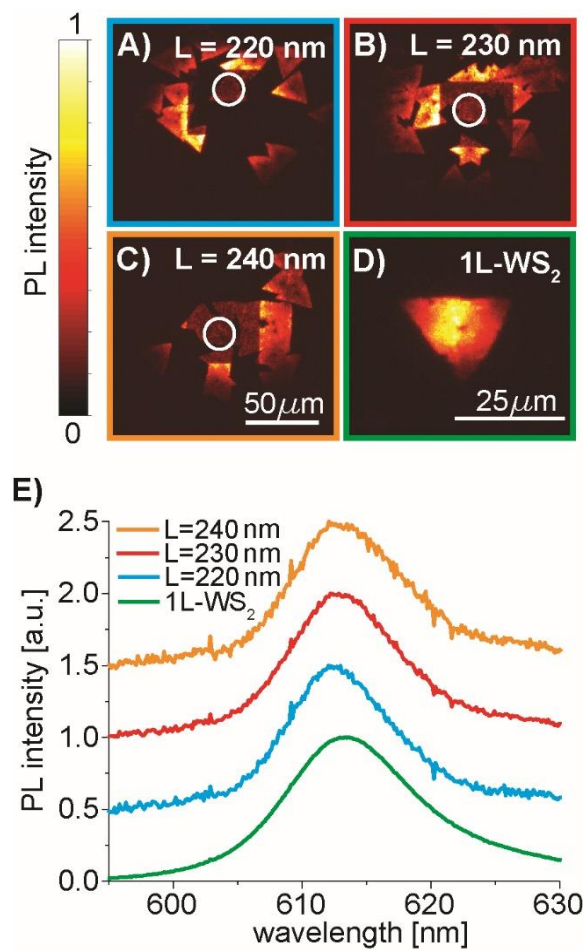

Figure 2: **Photoluminescence measurements for excitation at 532 nm wavelength.** (A-C) photoluminescence (PL) images of the areas around three different hybrid metasurfaces with different antenna lengths $L = 220 - 240$ nm. The white circles indicate the area, where the corresponding PL spectra are taken. The WS$_2$ flakes are clearly visible. The higher PL intensities close to the center of the image result from the excitation with an enlarged Gaussian beam shape (diameter ~50 µm) at 532 nm. (D) PL image of a single 1L-WS$_2$ flake without plasmonic nanorod antennas on top for comparison. (E) Corresponding PL spectra showing a clear peak at 614 nm. The spectra are shifted upwards by 0.5 for better visibility.

## 3. Nonlinear optical characterization

To characterize the interaction of the plasmonic nanorod antennas with the 1L-WS$_2$, we image the SHG signal of the samples in real-space by an optical microscopy setup (Figure 3A). As fundamental pump light, we use a pulsed laser beam at wavelengths between 1210 nm and 1270 nm with a pulse duration of 57 fs and 1 MHz repetition rate generated by an optical parametric amplifier system. The beam passes a quarter-wave plate, where it gets right circularly polarized ($\sigma^+$), is focused to a spot size of ~50 µm in diameter (FWHM) and illuminates the hybrid metasurface perpendicular. The emitted SHG light is then collected by a microscope objective in transmission and passed through a polarization analyzer to filter either the co- ($\sigma^+$) or cross-polarization state ($\sigma^-$) in respect to the fundamental beam before it is imaged onto an sCMOS camera.

Due to the scattering properties of the nanorod antennas and the coupling of them to the 1L-WS$_2$, the SHG light should contain both circular polarization states, as schematically illustrated in Figure 1A. In a simplified picture, the polarization dependency in such hybrid metasurfaces can be understood by splitting the entire process into two parts, the linear optical scattering at the Au-metasurface and the subsequent SHG of the scattered light at the 1L-WS$_2$. We note that in this picture the near-field effects of the plasmonic nanorod antennas are included and not treated separately.

As the circularly polarized incident beam excites localized surface plasmon polariton resonances in the nanorod antennas of the metasurface, the near-field and scattered light can be decomposed into two optical fields of opposite circular polarizations. We note that the field with the same circular polarization state as the incident beam does not carry any additional phase information while the opposite circular polarization state accumulates a Pancharatnam-Berry (PB) phase, which is determined by the orientation angle of the nanorod antenna [24, 25]. These two fields interact with the 1L-WS$_2$ to generate an SHG signal based on the high second-order nonlinearity of the 1L-WS$_2$. Due to the in-plane C$_3$ rotational symmetry of the hexagonal WS$_2$ lattice, the SHG is generated in the opposite polarization state [9, 26, 27]. Therefore, the helicity of the two scattered fields changes by the SHG process and only the resulting SHG field with the same polarization state (co-polarization) as the incident fundamental field carries the PB phase. Hence, the hybrid metasurface generates two circularly polarized SHG beams in different polarization states, one without an additional phase modulation in the cross-polarization $\sigma^-$ state and the other one with an additional PB phase modulation in co-polarization state $\sigma^+$ [10, 14, 19]. First, we study the SHG signals without utilizing the PB phase and focus solely on the polarization behavior of the hybrid metasurface.

### 3.1. Polarization dependency

To characterize the polarization state of the SHG for the hybrid metasurface, we image the emitted light of the sample surface by a home-build microscopy setup to a camera (Figure 3A). The obtained results for the two circular polarization states of the SHG signal and three different nanorod antenna lengths are shown in Figures 3B-D. The SHG signal of the samples in the cross-polarization state $\sigma^-$ is dominated by 1L-WS$_2$ without the metasurface on top. The corresponding SHG signal originates from the direct second-order nonlinear response of the 1L-WS$_2$ which covers the area around the hybrid metasurface. Based on the selection rules for circularly polarized light, this SHG light is generated solely in the cross-polarization state at the K and K' point of the WS$_2$. Hence, no SHG light is visible in the co-polarization state from the pure 1L-WS$_2$ flakes (see the bottom row in Figure 3B-D).

The observed lower SHG intensity in cross-polarization from the hybrid metasurface area is a result of the coupling of the light to the plasmonic nanorod antennas. Due to the plasmonic resonances, part of the incident fundamental light is reflected and absorbed, which reduces the total field strength at the WS$_2$ layer. The SHG intensity from the 1L-WS$_2$ below the Au-metasurface has mainly three contributions. First, the direct excitation by the fundamental beam that is not interacting with the nanorod antennas can result in SHG of the cross-polarization state $\sigma^-$. Second, the co-polarization near-field of the excited plasmonic resonances (which does not carry any PB phase) leads by the coupling to the WS$_2$ also to the emission of SHG in the cross-polarization state $\sigma^-$. Third, the cross-polarized near-field of the antennas that carries the PB phase with respect to the incident beam and generates SHG in the co-polarization state $\sigma^+$. Note that the SHG intensity is reduced closer to the border of the image due to the Gaussian beam shape and the finite beam size of the fundamental beam at 1240 nm wavelength.

The effect of the plasmonic metasurface on the SHG response is clearly visible if the same flake is divided into two parts, one with and the other without Au-nanorods on top. An example of such a flake, which is partially covered by the metasurface, is shown in Figure 3C. The edge of the Au-metasurface is clearly visible for the marked flake, which is covers approximately by half of the metasurface (left side of the flake). The magnified view in Figure 3E with the marked areas 1 and 2 show all four possible configurations. A strong SHG signal of the bare 1L-WS$_2$ without the metasurface is visible only for the cross-polarization state σ⁻ (blue marked area 2) confirming the selection rules for SHG at the 1L-WS$_2$. On the other hand, the SHG for the hybrid metasurface (green marked area 1) is visible in both polarization states as the nanoantennas scatter light in both circular polarization states of the fundamental light, which subsequently is frequency-converted by the WS$_2$.

While the hybrid metasurface provides a clearly visible SHG signal in both circular polarization states, the plasmonic nanoantennas themselves without the 1L-WS$_2$ below do not provide a significant SHG signal. The SHG intensity of solely Au-nanorods is close to the background signal from the bare glass substrate, as the intensities in both cases are nearly zero. This circumstance is clearly visible for both polarization states σ⁺ and σ⁻ of the SHG. The lack of an SHG signal of the Au-nanorod antennas is the result of their symmetry. While the plasmonic metasurface scatters light into both polarization states σ⁺ and σ⁻, almost no light is frequency-converted by a second-order nonlinear process in any polarization state, which is forbidden by the selection rules [19]. The remaining small signal only comes from the surface defects and the deviation from the perfect C2 symmetry.

Since the resonance wavelength of Au-nanorods changes by varying the length of the nanorods, we also investigate the influence of the nanorod length on the SHG signal of the hybrid metasurface. From Figures 3B-D the SHG signal strength can be determined for different antenna lengths of $L = 220 - 240$ nm. The SHG of the hybrid metasurfaces measured in co-polarization state $\sigma^+$ shows nearly the same SHG intensity for varying antenna lengths from 220 nm to 240 nm. Therefore, the different lengths of the Au-nanorods, designed in the different plasmonic metasurfaces, have only a weak influence on the SHG signal of the hybrid metasurface in co-polarization state $\sigma^+$.

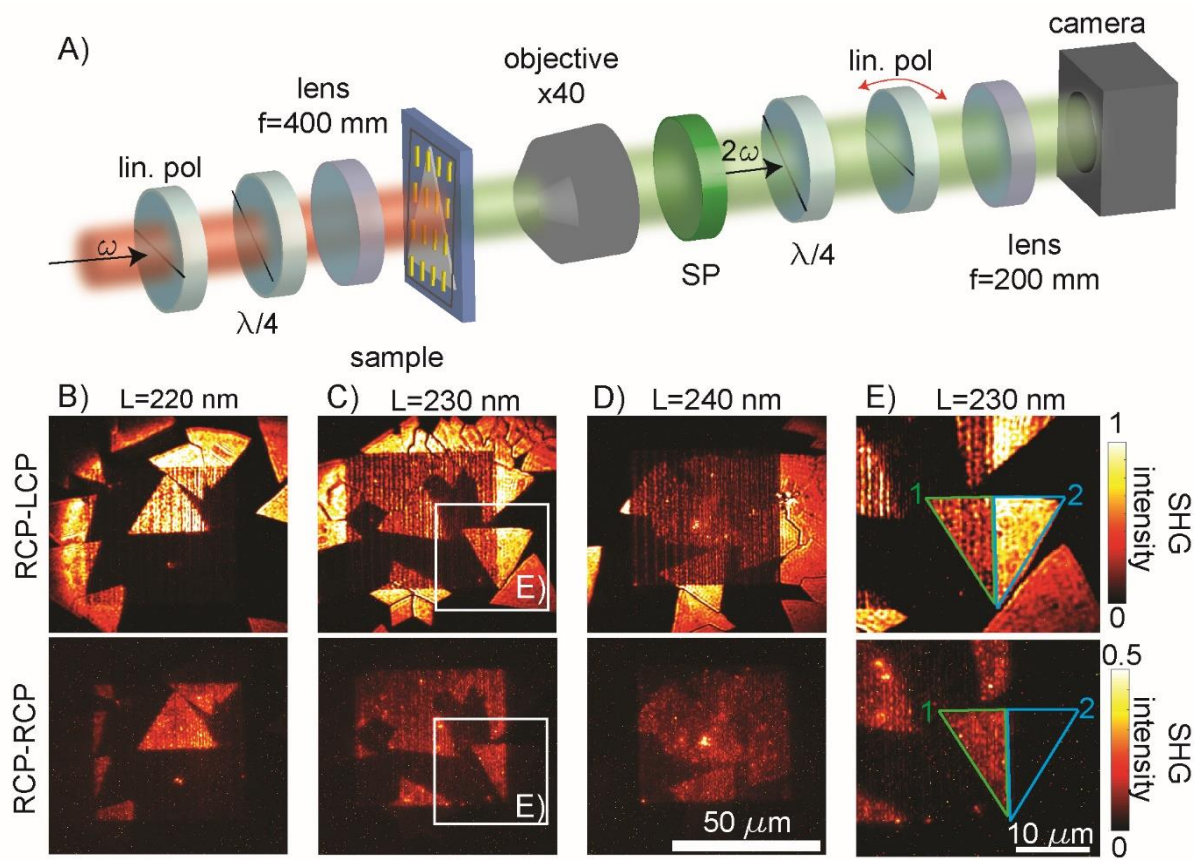

**Figure 3: Nonlinear optical measurement for the hybrid metasurfaces.** (A) Measurement setup for the spatially resolved SHG signal including polarization optics (analyzer). The fundamental laser beam is weakly focused on the sample and the polarization state is set by the linear polarizer and a quarter-wave plate. (B-D) SHG images of three different metasurfaces with antenna lengths of L=220-240 nm measured for the cross-polarization state $\sigma^-$ (LCP) (top row) and in co-polarization state $\sigma^+$ (RCP) (bottom row). The SHG signal of the surrounding 1L-WS$_2$ nearly vanishes in the co-polarization state $\sigma^+$ due to the selection rules for the SHG process. (E) Close-up of the marked area (white box) in (C) showing the edge of the metasurface passing through a 1L-WS$_2$ flake. The area 1 (marked in green) points out an area covered with Au-nanoantennas while the area 2 (marked in blue) points out the same flake which is not covered with Au-nanoantennas. The SHG signal in the blue marked area 2 vanishes for measurement in the co-polarization state.

### 3.2. Wavelength dependency

The SHG efficiency from the 1L-WS$_2$ strongly depends on the fundamental wavelength. For wavelengths close to half the bandgap energy, the nonlinear process becomes strongly enhanced [28]. In addition, the plasmonic resonances of the nanoantennas can further enhance the SHG process depending on the spectral location of their resonances [29-31]. However, compared to the band edge emission of the WS$_2$ the plasmon resonances of the Au-antennas are much broader so that for a slight shift of the resonance frequency might not influence the efficiency strongly. For our study of the wavelength dependency, we chose the metasurface with the antenna length of 230 nm, which gives rise to a plasmon resonance at 1200 nm. Figure 4 shows the measured spatially resolved SHG signals for different fundamental wavelengths for both polarization states of the SHG.

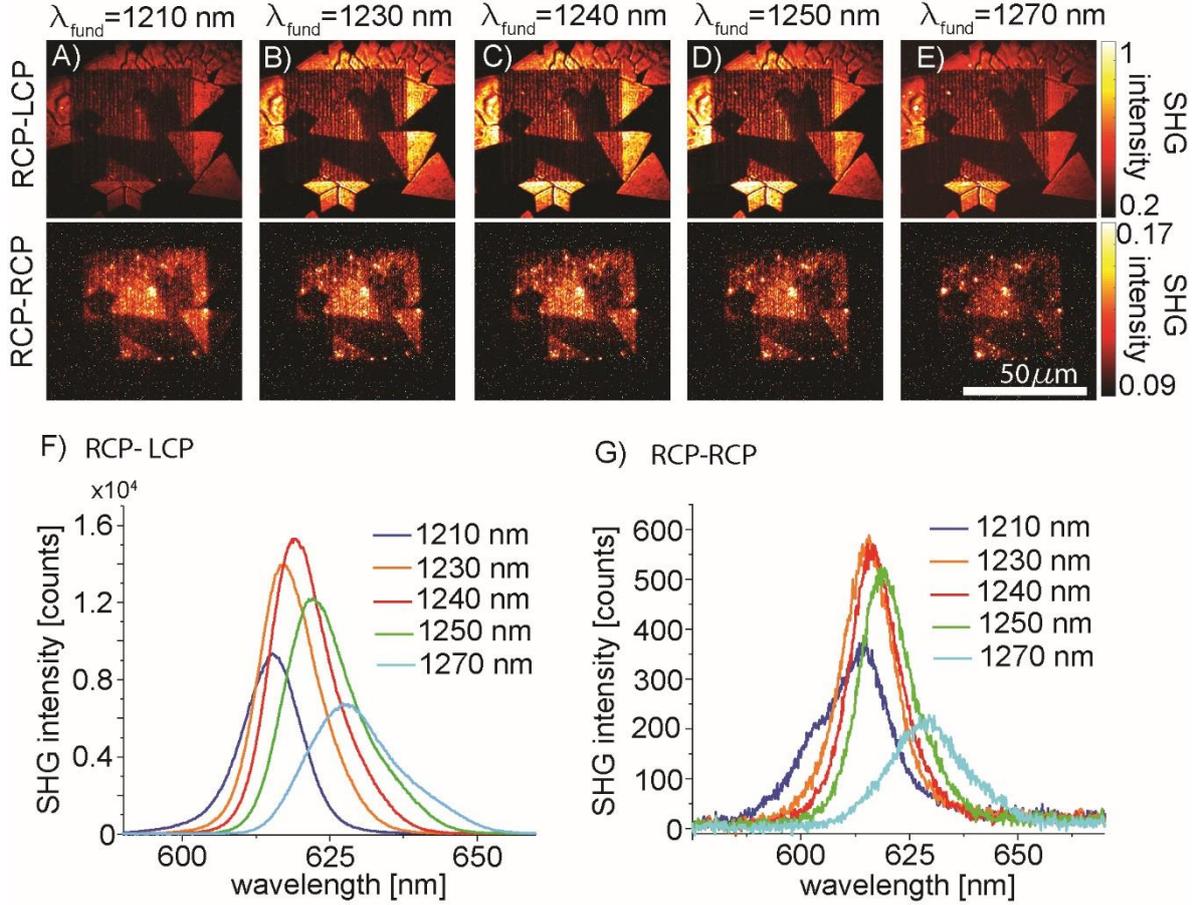

**Figure 4: Wavelength dependence of the hybrid metasurface.** (A-E) Wavelength dependent measurement in the range of $\lambda_{\text{fund}} = 1210 - 1270$ nm measured in the cross-polarization state $\sigma^-$ (RCP-LCP) and the co-polarization state $\sigma^+$ (RCP-RCP). (F) Obtained SHG of the hybrid metasurfaces spectra measured in the cross-polarization state $\sigma^-$. The strongest SHG is observed for a fundamental wavelength of 1240 nm. (G) Obtained spectra measured of the hybrid metasurfaces in the co-polarization state $\sigma^+$. The curves show nearly the same wavelength dependence as for the cross-polarization state in (G) despite the lower intensity.

The strongest SHG signal for the bare 1L-WS$_2$ (see areas around the hybrid metasurface) is expected for a wavelength of half the bandgap energy, which is for the 1L-WS$_2$ at $\frac{1}{2}E_{\text{gap}} \approx 1$ eV $\triangleq 1240$ nm. The spatially resolved SHG images of the hybrid metasurface show the strongest SHG signal measured in cross-polarization state $\sigma^-$ appears at a fundamental wavelength of 1240 nm (Figure 4F). For that polarization, the SHG signal originates mostly from bare 1L-WS$_2$ and is therefore influenced by the bandgap energy.

For the co-polarization state $\sigma^+$, only the hybrid metasurface contributes to the SHG process. Since the Au-nanorods have a resonance wavelength at $\lambda \approx 1200$ nm, this can result in a shift for the highest SHG efficiency. Figure 4G shows for the hybrid system that the SHG signal reaches its maximum at a fundamental wavelength of $\lambda_{\text{fund}} = 1230$ nm, which is slightly blue-shifted compared to the bare WS$_2$ flakes. As shown before, the influence of the resonance of the Au-nanorods, determined by the length, has only a weak impact on the SHG signal of the hybrid metasurface in co-polarization state $\sigma^+$. To obtain a greater SHG signal, one might optimize the length of the Au-nanorods to achieve a better overlap among the plasmon resonance and the SHG resonance of the 1L-WS$_2$. The overall SHG efficiency for the hybrid metasurface for SHG generation in the co-polarization state $\sigma^+$ is calculated to ~$10^{-13}$ for a fundamental wavelength of 1240 nm.

Due to the strong pump beam at half of the band gap energy, potentially two-photon absorption can occur, resulting in the emission of PL by the recombining excitons. However, the SHG is the dominating process and therefore the response of the hybrid metasurface in the nonlinear case is instantaneous. This can be seen from the spectral width (FWHM) of the SHG (Figure 4F), which is a convolution of the laser spectrum and the instantaneous spectral nonlinear response function. On the other hand, the PL spectrum has only half of the spectral width as for the SHG spectrum (Figure 2E), because it is only determined by the exciton lifetime and the inhomogeneous broadening. Furthermore, the strong excitation with the laser pulses can result in saturation effects of the SHG process. To confirm that there are no saturation effects by the strong pump field, we performed a power dependent measurement of the SHG signal, which show almost no deviation from the quadratic dependence (see supplementary material).

### 3.3. Spatial phase control of the SHG

Next, we show the potential of the hybrid metasurface for further control of the SHG by utilizing the geometric phase provided by the nanorod antennas (see Figure 1A). Therefore, we introduce a space-variant Pancharatnam-Berry (PB) phase [32-34] by means of the plasmonic nanorod orientation in order to control the radiated SHG intensity in the far-field [24]. As stated before, a circularly polarized incident beam excites a localized surface plasmon polariton resonance in the nanorod antennas with a strong dipole moment along the long axis of the nanorods. The near-field and scattered light by these induced dipoles can be decomposed into two optical fields of opposite circular polarizations. We note, that the scattered light in the cross-polarization state $\sigma^-$ carries additional phase information, which depends on the local orientation of the nanorods. In the linear optical case, the local phase $\varphi$ of the transmitted fundamental light is determined by the rotation angle $\theta$ of the individual nanorods by $\varphi = 2\theta$ [25].

We now focus on the transmitted fundamental beam in the cross-polarization state $\sigma^-$, which carries the PB phase information and interacts with the 1L-WS$_2$. The subsequent nonlinear SHG process alters the PB phase induced by the Au-nanorods by adding another factor of two resulting in a PB phase of the SHG light by $\varphi = 4\theta$, which is now four-times the rotation angle of the nanorods [14, 19]. We note, that the phase carrying SHG signal is now encoded in the co-polarization state $\sigma^+$, due to the polarization changing SHG process in the 1L-WS$_2$.

In order to visualize this additional PB phase in the co-polarization state of the SHG signal, we introduce a spatial phase modulation over the entire hybrid metasurface. For that, we choose two metasurfaces with two different spatially separated areas of local phase modulations to demonstrate a spatial interference effect of the SHG in the far-field [35].

Figure 5A shows the three different arrangements of the nanoantennas, marked in green (1), purple (2) and blue (3) to generate various interference effects. In area 1 (green) all nanorods have the same orientation angle ($\theta = 0°$) and generate a local SHG signal carrying the same PB phase. For this kind of phased-array of nanorods, the SHG signal interferes constructively in forward direction in the far-field. In area 2 (purple) every second row consists of nanorods rotated by an angle of $\theta = 45°$. The resulting phase shift of the SHG signal from the alternating rows is then $\varphi = 180°$ with respect to the non-rotated nanorods. Hence, the SHG signals from these two rows interfere destructively in the far-field, resulting in a reduced SHG signal compared to the SHG signal resulting from area 1. In area 3 (blue) every second row consists of nanorods rotated by an angle of $\theta = 90°$. The phase shift of the SHG signals between these two rows is then $\varphi = 4\theta = 360°$ and therefore the SHG signal again interferes constructively in forward direction in the far-field.

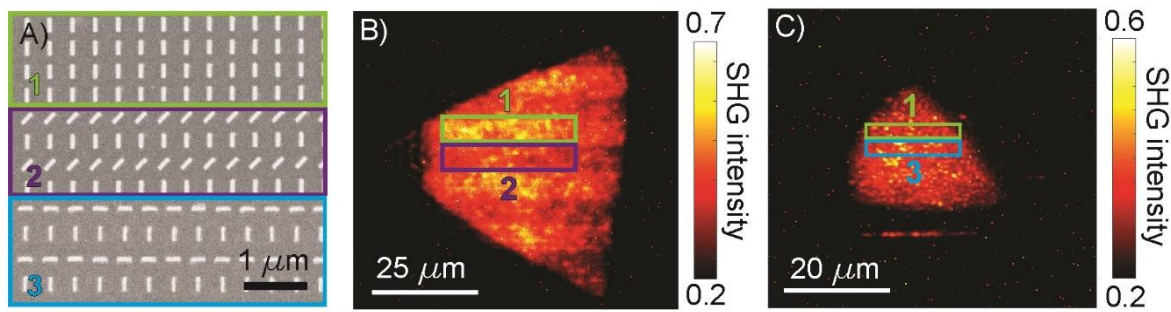

Figure 5: Spatial phase control of the SHG signal. (A) SEM image of the three distinct areas to generate different local phase distributions. In area 1 (marked in green) the plasmonic nanorods all have the same orientation. In area 2 (marked in purple) half of the antennas have the same orientation as the antennas in area 1. The other half (every second row) is rotated by an angle of $\theta = 45°$ resulting in an additional PB phase of $\varphi = 180°$. In area 3 (blue) the previously 45° rotated antennas are now rotated by $\theta = 90°$ resulting in an additional PB phase of $\varphi = 360°$ (B) SHG image of the plasmonic metasurface with alternating areas 1 and 2 on top of a 1L-$WS_2$ flake measured in co-polarization state $\sigma^+$ at $\lambda = 620$ nm. The marked areas (green and purple) correspond to the areas with constructive and destructive interference, respectively. (C) SHG image of the plasmonic metasurface with areas 1 and 3 on top of a 1L-WS2 flake measured again in co-polarization state $\sigma^+$. The marked areas (green and blue) indicate the corresponding areas with plasmonic nanorods.

To investigate the influence of an implemented PB phase distribution in the plasmonic metasurface, we first image the SHG signal of the hybrid metasurface with alternating areas 1 and 2 as shown in Figure 5A in the co-polarization state $\sigma^+$. For the SHG image in cross-polarization state $\sigma^-$ see the supplementary material. The SHG image of the hybrid metasurface with the implemented antenna distributions is shown in Figure 5B and shows a clear stripe pattern in the SHG signal. The comparison between the areas 1 and 2 shows an intensity difference of the SHG signal although all nanorod antennas have the same properties. For the destructive case, the SHG intensity drops by a factor of nearly 2 compared to the areas with constructive interference. This leads to the conclusion, that the local phase manipulations by the Au-nanorods can be transferred to the SHG signal originating from the 1L-WS$_2$. However, the SHG signal is not completely suppressed in area 2, which leads to the conclusion that the near-field coupling effect might also generate other phase components.

To further investigate the spatial phase modulation, we fabricated another hybrid metasurface with a different spatial phase modulation pattern. The metasurface now consists of nanorod arrays as shown in the areas 1 and 3 (Figure 5A). The resulting spatial phase introduced from every second row of the nanorods in area 3 in the co-polarization state σ$^+$ is $\varphi = 4\theta = 360°$ with respect to the non-rotated nanorods. Therefore, all nanoantennas interfere constructively despite their rotation and no spatial intensity modulation in the far-field is expected for the entire hybrid metasurface. The corresponding SHG image in the co-polarization state σ$^+$ is shown in Figure 5C, while the SHG image in the cross-polarization state σ$^-$ can be found in the supplementary material. As a reference, area 1 (green) and 3 (blue) mark the position of areas with different antenna patterns, shown in Figure 5A, within the hybrid metasurface exemplary. As expected, we do not observe any kind of intensity modulation in the far-field in the co-polarization state σ$^+$ and a homogeneous intensity pattern is visible.

## 4. Conclusions

In summary, we present experimental results of the nonlinear coupling in a hybrid metasurface consisting of plasmonic metasurfaces on top of a monolayer WS$_2$. After the characterization of the linear optical properties of hybrid metasurfaces, we analyze the impact of different parameters, e.g. polarization and wavelength, on the coupling by imaging spatially resolved the SHG signal. For such hybrid metasurfaces, we observe an SHG signal co-polarization state (σ$^+$), which is forbidden for either solely 1L-WS$_2$ or Au-nanorods by process conditioned selection rules. Polarization-dependent measurements show the spatial SHG intensities for different compositions (hybrid metasurface, solely 1L-WS$_2$, bare Au-metasurface). In addition, the strength of the SHG signal in the hybrid metasurface depends on different parameters. While a slight change of the length of the Au-nanorods has only a weak impact, the exciting wavelength affects the SHG signal to a greater extent. The resonance wavelength of the hybrid metasurface is defined by the bandgap of the 1L-WS$_2$ and the resonance condition of the Au-nanorods. Hybrid metasurfaces can be used to overcome the limitations of two process conditioned selections rules to generate a second harmonic signal which would be not allowed by either of them. To show the potential of a hybrid metasurface for further applications, we added additional phase information to the SHG signal by inducing a local Pancharatnam-Berry phase due to the interaction with the nanorod antennas with the fundamental incident beam. In such way, we demonstrate an additional degree of freedom for tailoring the far-field intensity of the SHG signal resulting from the hybrid metasurface.

## Acknowledgments

The authors would like to acknowledge the continuous support from Cedrik Meier by providing access to the electron beam lithography system. This project has received funding from the European Research Council (ERC) under the European Union's Horizon 2020 research and innovation programme (grant agreement No 724306), the NSFC-DFG joint research program (DFG ZE953/11-1, NSFC No. 61861136010), and the National Natural Science Foundation of China (No. 61775019).